\documentclass[%
 reprint,
 amsmath,amssymb,
 aps,
]{revtex4-2}

\usepackage{graphicx}
\usepackage{dcolumn}
\usepackage{bm}
\usepackage[mathlines]{lineno}
\usepackage{lineno}
\usepackage{float}

\begin{document}

\preprint{APS/123-QED}

\title{Characterization of resonator using confocal laser scanning microscopy and its
application in air density sensing}

\author{Ayla Hazrathosseini}
 \affiliation{Department of Physics and Astronomy, Texas A\&M University, College Station, TX 77843, USA\\
 Institute of Quantum Science and Engineering, Texas A\&M University, College Station, TX  77843, USA}
 
\author{Mohit Khurana}%
 \email{mohitkhurana@tamu.edu \\ AH and MK contributed equally to this work.}
\affiliation{Department of Physics and Astronomy, Texas A\&M University, College Station, TX 77843, USA\\
 Institute of Quantum Science and Engineering, Texas A\&M University, College Station, TX  77843, USA}%

\author{Lanyin Luo}
\affiliation{Department of Physics and Astronomy, Texas A\&M University, College Station, TX 77843, USA\\
Institute of Quantum Science and Engineering, Texas A\&M University, College Station, TX  77843, USA}%

\author{Zhenhuan Yi}
\affiliation{Department of Physics and Astronomy, Texas A\&M University, College Station, TX 77843, USA\\
 Institute of Quantum Science and Engineering, Texas A\&M University, College Station, TX  77843, USA}%

\author{Alexei Sokolov}
\affiliation{Department of Physics and Astronomy, Texas A\&M University, College Station, TX 77843, USA\\
 Institute of Quantum Science and Engineering, Texas A\&M University, College Station, TX  77843, USA}%

\author{Philip R. Hemmer}
\affiliation{Department of Physics and Astronomy, Texas A\&M University, College Station, TX 77843, USA\\
Institute of Quantum Science and Engineering, Texas A\&M University, College Station, TX  77843, USA\\
Department of Electrical and Computer Engineering, Texas A\&M University, College Station, TX 77843, USA}%

\author{Marlan O. Scully}
\affiliation{Department of Physics and Astronomy, Texas A\&M University, College Station, TX 77843, USA\\
 Institute of Quantum Science and Engineering, Texas A\&M University, College Station, TX  77843, USA\\
 Baylor University, Waco, Texas 76704, USA\\
 Princeton University, Princeton, New Jersey 08544, USA}%

\begin{abstract}We present the characterization of the photonic waveguide resonator using confocal laser scanning microscopy imaging method. Free space TEM$_{00}$ laser mode is coupled into quasi-TE$_{0}$ waveguide mode using confocal microscopy via a diffractive grating coupler and vice versa. Our work includes the design, fabrication, and experimental characterization of a silicon nitride racetrack-shaped resonator of length $\sim 165\,\mu m$. We illustrate clear evidence of resonance excitation from the confocal microscope image and demonstrate loaded Q-factor and finesse $\sim 8.2\pm 0.17 \times10^{4} $ and $\sim 180\pm 3.5$, respectively. We further demonstrate its one application in air density sensing by measuring the resonance wavelength shifts with variation in environment air pressure. Our work impacts spectroscopy, imaging, and sensing applications of single or ensemble atoms or molecules coupled to photonic devices. Additionally, our study highlights the potential of confocal microscopy for analyzing photonic components on large-scale integrated circuits, providing high-resolution imaging and spectral characterization.

\end{abstract}
\maketitle
\section{Introduction}
\noindent 
Confocal microscopy offers significant advantages over traditional camera-based imaging 
and techniques for investigating and characterizing quantum emitters, photonic waveguides, and cavities \cite{Elliott2020-lz, Akimov_2007, Kulkarni2008, AbouKhalil2017, Rugar2020-ge}. Confocal microscopy significantly enhances photon collection precision by focusing on a small spot on the sample and using a pinhole to filter out-of-focus light, thereby reducing background noise. This technique ensures that only photons from the targeted, localized region are collected, which is crucial for applications requiring high spatial selectivity, such as single-particle dynamics probing. In contrast, standard microscopy collects light from the entire field of view, including out-of-focus regions, leading to higher background signals that can overwhelm the desired signal. As a result, while standard microscopy is suitable for general imaging, it lacks the ability to isolate photons from specific, small regions, making confocal microscopy the superior choice for applications demanding precise photon collection and superior accuracy. Its superior spatial resolution and optical sectioning capability allow for detailed examination of individual photonic elements, even in densely packed circuits \cite{Pawley}. Rejecting out-of-focus light enhances contrast and enables precise measurements of device geometries and surface qualities \cite{WILSON_2011}. Moreover, confocal microscopy's depth discrimination facilitates three-dimensional imaging, which is crucial for analyzing multi-layer structures or vertical coupling regions \cite{Novotny_2006}. The technique's capacity for spectral analysis can provide insights into the optical properties and performance of each resonator or cavity coupled to particles \cite{Becker_2003}. 

\noindent 
Confocal microscopy with advanced computation technique allowed deep-label free imaging \cite{Chen2023}. On the other hand, label-free detection of nano-scale objects such as nanoparticles and various biological materials (cell, virus, protein, DNA, etc.) coupled to whispering gallery modes (WGM) and microresonators have been explored \cite{Vollmer_2008, VollmerArnold2008}. 
Kim and coworkers reported size measurement of single nanoparticles in aquatic environments using mode splitting in optical resonators and co-related particle size and resonator Q-factor \cite{kim_2011, kimandcoworkers_2012}. Upconversion particles suspended in a solution can be effectively detected and analyzed using confocal microscopy, which can allow tracking of their movement with high spatial resolution. These particles, which convert lower-energy photons into higher-energy ones, can be visualized in real-time by capturing their motion as a movie using reflection and fluorescence imaging and spectroscopy through confocal microscopy. Integrating these upconversion particles with an optical resonator can add further advantages in sensing capabilities \cite{Ren_2022, Sahoo_2022}. The optical resonator enhances the interaction between light and the particles, amplifying the upconverted signal and enabling sensitive detection of changes in the particles' environment. This proposed approach would combine the precise imaging of confocal microscopy with the enhanced sensitivity of optical resonators, making it a powerful technique for advanced spectroscopy and sensing applications. However, the characterization of resonators with sensing capabilities using confocal microscopy has yet to be demonstrated. This motivates us to apply the confocal microscopy method; otherwise, new possibilities would remain undiscovered with any other imaging technique.

\noindent
In this work, we implement a waveguide-coupled racetrack-shaped silicon nitride ($\text{Si}_{3}\text{N}_{4}$) resonator on silica photonic platform to demonstrate its characterization using a two-channel confocal laser scanning microscopy. Note that (to the best of our knowledge), this is the first time when the radiated photons from a small part of the microresonator are collected to a fiber. We perform air density sensing measurements matching the theoretical estimations, which further validate the robustness of the applied method.  The sensing capability achieved in this work is better than the works reported in \cite{Singh2019, Elmanova2021, Antonacci2020, Feng2015, Kazanskiy2023, Robinson2008}. We expect our work to inspire new developments in the exploration of cavity QED and optomechanics using confocal laser scanning microscopy imaging techniques.

\subsection{Characterization and confocal imaging of photonic devices}
\noindent
Different mechanisms of coupling light into a waveguide, such as laser-diode or fiber to waveguide (e.g., butt-coupling, tapered fiber to tapered waveguide \cite{Alajlan2020-aj, Khurana2022-as}), laser-diode or fiber to grating coupling and confocal microscopy (as presented in this work), offer distinct advantages and are suited to specific applications. The fiber-to-waveguide coupling method directly connects an optical fiber to a photonic chip, typically through edge coupling, where the fiber is aligned to a chip waveguide for maximum mode overlap. This method offers high efficiency but requires precise alignment and is sensitive to position, making it challenging to maintain stability over time. The fiber-to-grating coupling method uses a grating etched onto the chip's surface to couple light from a fiber into the waveguide. This method allows for easier alignment and can accommodate slight misalignment better than the fiber-to-waveguide methods, but it typically suffers from lower efficiency due to diffraction losses. The confocal microscopy method (as presented in this work) allows us to couple light into and out from the grating structures in small regions of the sample with a stable optical setup, requires no hard-to-implement light-to-chip coupling scheme, and is robust in comprehending scattering points and light-matter dynamics in micro-sized photonic devices with much higher resolution which is impossible with the gold standard mechanisms. The optical setup used in the demonstration experiment of this work also gives an edge in characterizing and operating a photonic integrated circuit in ambient and cryogenic conditions. Although confocal microscopy does not offer high coupling efficiency compared to the fiber-chip waveguide butt-coupling method, its advantages are more far-reaching than any other technique. Moreover, the coupling efficiency can be improved by inverse-designed structures specifically designed for confocal microscopy characterization method using adjoint optimizations substituting the grating couplers, a subject that goes beyond the scope of this work. 

\noindent
In the wider context of our work's impact, we can use automated scanning and image processing to characterize multiple photonic elements efficiently within a large-scale photonic circuit. This would enable us to address the challenge of assessing individual devices. The combination of high-resolution imaging, 3D capability, and potential for automation makes confocal microscopy an invaluable tool for comprehensive and efficient characterization of complex photonic integrated circuits, offering a level of detail and throughput that is difficult to achieve with conventional imaging methods \cite{Archetti_2019, Holler_2017}.

\subsection{Air density sensing}
\noindent Ideal gas law, $\text{PV = nRT = NK}_{\text{B}}\text{T}$, where P is pressure, V is volume, n is number of moles, N is number of particles, R is Universal gas constant, $\text{K}_{\text{B}}$ is Boltzmann constant, T is temperature dictates the behavior of air particles in environment. The pressure is directly related to the number of particles present in a given volume. Resonators are highly precise optical instruments that leverage high-Q resonance to detect minute disturbances and accurately measure changes in the refractive index of air or the material being studied. The operating principle of resonators is based on the resonance shift caused by alterations in the refractive index of the surrounding medium, such as air, or changes in the number of particles in the medium. When light is coupled with the resonator, it propagates through the waveguide, interacting with the resonator's material and the surrounding medium. The resonant wavelength of the resonator is determined by the resonator's circumference and the effective refractive index of the waveguide material, which is influenced by the refractive index of the surrounding medium \cite{Toropov2021}. Changes in the refractive index of the surrounding medium impact the effective index ($\text{n}_{\text{eff}}$) of the mode, resulting in a shift in the resonant wavelength of the resonator. Similarly, changes in the number of particles in the medium, which affect its effective index, lead to a shift in the resonant wavelength of the resonator. By monitoring this shift in the resonant wavelength, the resonator can be used to measure changes in the refractive index of air or the number of particles in the air with high sensitivity, making it a valuable tool for a wide range of applications, from environmental monitoring to gas sensing \cite{Kazanskiy2023, Huang2022, Gao2016, Foreman_2015}. For an ideal gas with n $\approx$ 1, the refractive index is related to density using the Dale-Gladstone equation \cite{Wanstall2020},
$\text{n} - \text{1} = \text{K}\cdot\rho$, where n is the refractive index of air, K is the Gladstone-Dale constant, $\rho$ is the density of the medium. For small changes in pressure, the change in refractive index $(\Delta \text{n})$ can be approximated by, 
$\Delta \text{n} \approx \frac{\text{K}}{\text{R T}} \cdot \Delta \text{P}$,
where, $\Delta \text{n}$ is change in refractive index,
$\Delta \text{P}$ is the change in pressure, R is the specific gas constant for dry air, and T is the absolute temperature in Kelvin \cite{Wanstall2020}.

\noindent
The effective area spanned by an electromagnetic (EM) wave mode is given by $\text{A}_{\text{eff, mode}}$ \cite{Agrawal2019},

\begin{equation} \label{eq:2}
\text{A}_\text{eff, mode} = \frac{\iint_\text{mode} \text{n}^{2}\text{(x,y)}|\text{E(x,y)}|^{2} \,\text{dx}\,\text{dy}}{\max[\text{n}^{2}\text{(x,y)}|\text{E(x,y)}|^{2}]_\text{mode}}
\end{equation}

\noindent where, $\text{A}_\text{eff, mode}$ is the effective mode area, $|\text{E(x,y)}|$ is the electric field of mode and $\text{n(x,y)}$ is the refractive index. Therefore, the cross-sectional area of air particles spanned by an EM mode is given by $\text{A}_\text{eff, air}$,

\begin{equation}
\label{eq:4}
\text{A}_\text{eff, air} = \frac{\iint_\text{air} \text{n}^{2}\text{(x,y)}|\text{E(x,y)}|^{2} \,\text{dx}\,\text{dy}}{\max[\text{n}^{2}\text{(x,y)}|\text{E(x,y)}|^{2}]_\text{mode}}
\end{equation}

\noindent
The number of air particles in the sensing volume of an EM wave mode propagating in the resonator contributing to the wavelength shift is given by, $\frac{\text{PV}_{\text{sensing}}}{\text{K}_{\text{B}}\text{T}}$, where $\text{V}_{\text{sensing}}=\text{A}_{\text{eff, air}}\text{L}_{\text{resonator}}$. 

\section{Design and Simulations}
\begin{figure}[ht!]
\centering\includegraphics[width=8.5cm]{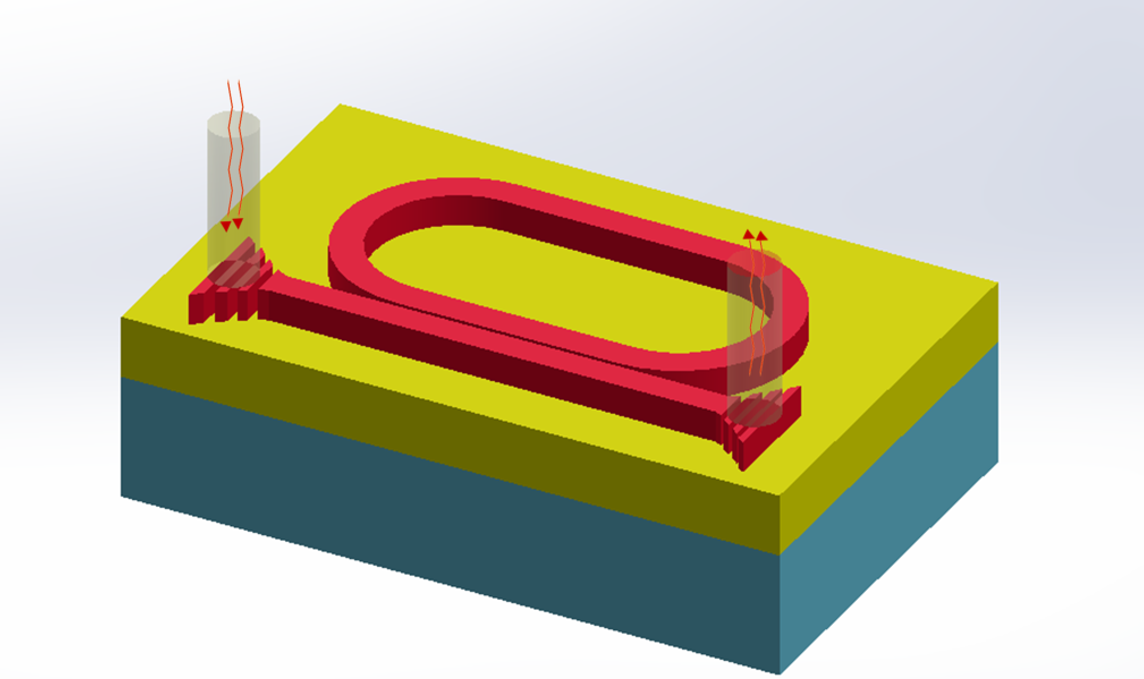}
\caption{Schematic diagram of $\text{Si}_{3}\text{N}_{4}$ waveguide coupled racetrack shaped resonator device, 200 nm thick $\text{Si}_{3}\text{N}_{4}$ sits on top of 2 $\mu$m thick $SiO_{2}$ cladding on Si substrate. Light is coupled to a waveguide via a grating coupler using the confocal laser scanning microscopy method. Light is out-coupled from the waveguide via a grating coupler and collected in a confocal microscope.}
\label{experimental_schematic_diag}
\end{figure}

\begin{figure}[ht!]
\centering\includegraphics[width=8.5cm]{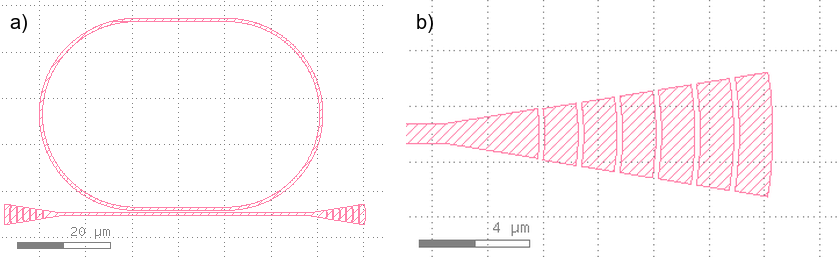}
\caption{Top view images from graphic design system viewer; a) a large view of waveguide coupled racetrack-shaped resonator and b) a close view of an optimized grating coupler used in the demonstration experiment.}
\label{GDS_image}
\end{figure}

\noindent 
Thin film $\text{Si}_{3}\text{N}_{4}$ resonators have emerged as versatile photonic components in integrated photonics, finding applications across a wide spectrum of fields. It offers low-propagation losses in waveguides, high Q-factors resonators, and complementary metal-oxide semiconductor (CMOS) compatibility, making them ideal for various cutting-edge applications \cite{Chauhan2022}. We implement $\text{Si}_{3}\text{N}_{4}$ on silicon dioxide ($\text{SiO}_{2}$) cladding photonic platform and utilize a two-channel confocal microscope for waveguide and resonator characterization. TEM$_{00}$ mode of laser is coupled into $\text{TE}_{0}$ mode of the waveguide via a grating coupler using one confocal channel (A). Light propagating in TE$_{0}$ waveguide mode is out-coupled via the second grating coupler to free-space TEM$_{00}$ mode and collected in the second confocal channel (B). The schematics diagram is shown in Fig. \ref{experimental_schematic_diag}. The grating coupler efficiently couples light from TEM$_{00}$ mode into a waveguide mode on a photonic chip. The TEM$_{00}$ mode, which is the lowest-order TEM mode, has a Gaussian intensity profile and is common in single-mode optical fibers and laser propagation (as in confocal microscopy setup). The challenge in coupling this mode into a waveguide lies in the difference in mode profiles and the need to match the momentum of the light focused by the objective with that in the waveguide. The grating coupler consists of periodic structures etched into the surface of the chip (see Fig. \ref{GDS_image}). When the TEM$_{00}$ mode light from the objective lens of the confocal microscope is incident on the grating, the periodic variation in the refractive index creates a diffraction effect. This diffraction effectively converts the freely propagating light into a mode that can be confined within the waveguide. The grating is designed to ensure that the light is diffracted at an angle that allows it to satisfy the phase-matching condition between the fiber $\text{TEM}_{00}$ mode and the waveguide TE$_{0}$ and TM$_{0}$ modes. The mode conversion process involves carefully designing the grating parameters, such as the grooves' period, depth, and shape, to match the effective index of the waveguide mode. The grating not only redirects the incoming light but also gradually transforms the spatial profile of the TEM$_{00}$ mode into the spatial profile of the waveguide mode. This allows the light to couple efficiently into the waveguide with minimal loss. Similarly, the light is out-coupled from waveguide mode to the free-space TEM mode and sent to the objective lens via a grating coupler.

\noindent
The resonant condition of a resonator is given by $\text{m}\lambda = \text{L}_{\text{eff}}$, where m is an integer (the mode number), $\lambda$ is the wavelength of the light in vacuum, $\text{L}_{\text{eff}}$ is the effective optical path length of the resonator. The free spectral range (FSR) of a resonator is given by $\frac{\lambda^2}{\text{n}_\text{g}\text{L}_{\text{eff}}}$, where $\text{n}_{\text{g}}$ is the group index of the waveguide, $\lambda$ is the operating wavelength. Group index is given by, $\text{n}_{\text{g}}=\text{n}_{\text{eff}}-\lambda\frac{\text{dn}_{\text{neff}}}{\text{d}\lambda}$, where $\text{n}_{\text{eff}}$ is effective mode index and dependent on wavelength. The quality factor (Q-factor) of resonance is evaluated by $\text{Q} = \frac{\lambda_0}{\Delta \lambda}$, where $\lambda_0$ is the central resonant wavelength and $\Delta \lambda$ is the full width at half maximum (FWHM) of the resonant peak.

\begin{figure}[ht!]
\centering\includegraphics[width=8.5cm]{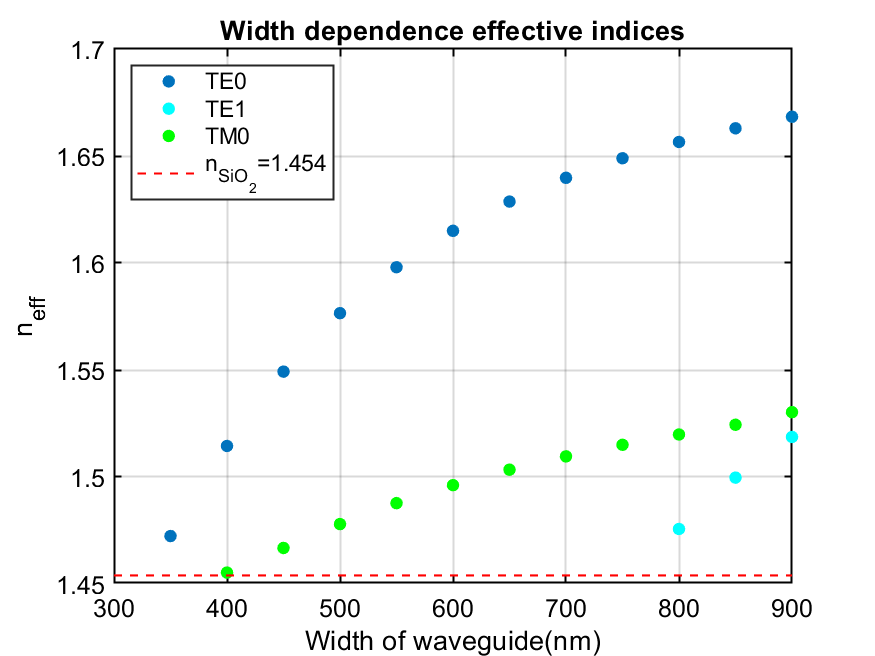}
\caption{Effective indices of modes confined in 200 nm thick varied width $\text{Si}_{3}\text{N}_{4}$ core waveguide on 2 $\mu\text{m}$ $\text{SiO}_{2}$ on Si substrate estimated at 772 nm. The refractive index of SiO$_{2}$ cladding is 1.454, which is plotted as a red dotted line. Any mode with an index below the cladding index is not a guided mode in the core waveguide.}
\label{neff_modes}
\end{figure}

\begin{figure}[ht!]
\centering\includegraphics[width=8.5cm]{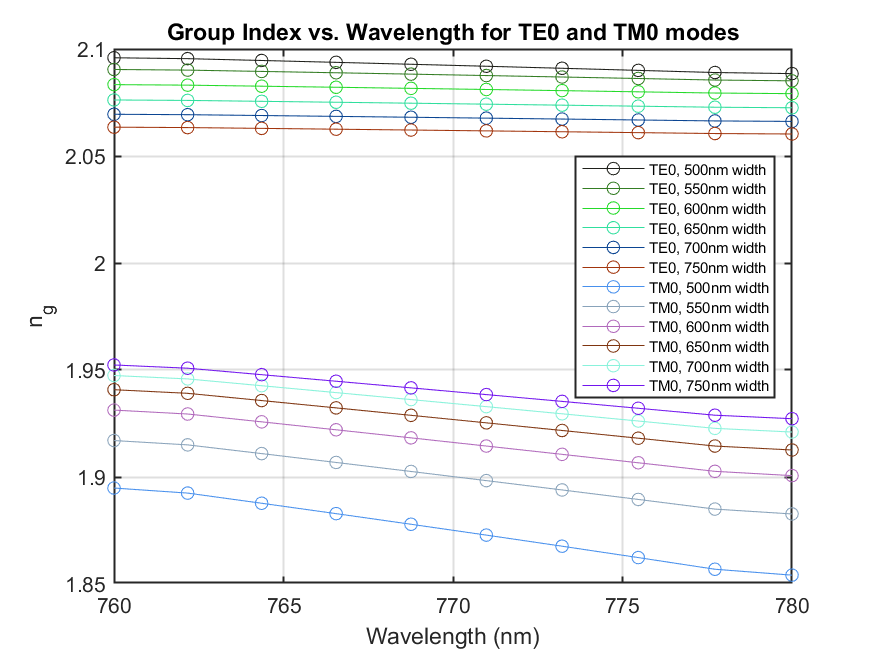}
\caption{Wavelength dependence group indices of $\text{TE}_{0}$ and $\text{TM}_{0}$ modes confined in 200 nm thick varied width $\text{Si}_{3}\text{N}_{4}$ waveguide on 2 $\mu\text{m}$ $\text{SiO}_{2}$ on Si substrate.}
\label{ng_all}
\end{figure}

\begin{figure}[ht!]
\centering\includegraphics[width=8.5cm]{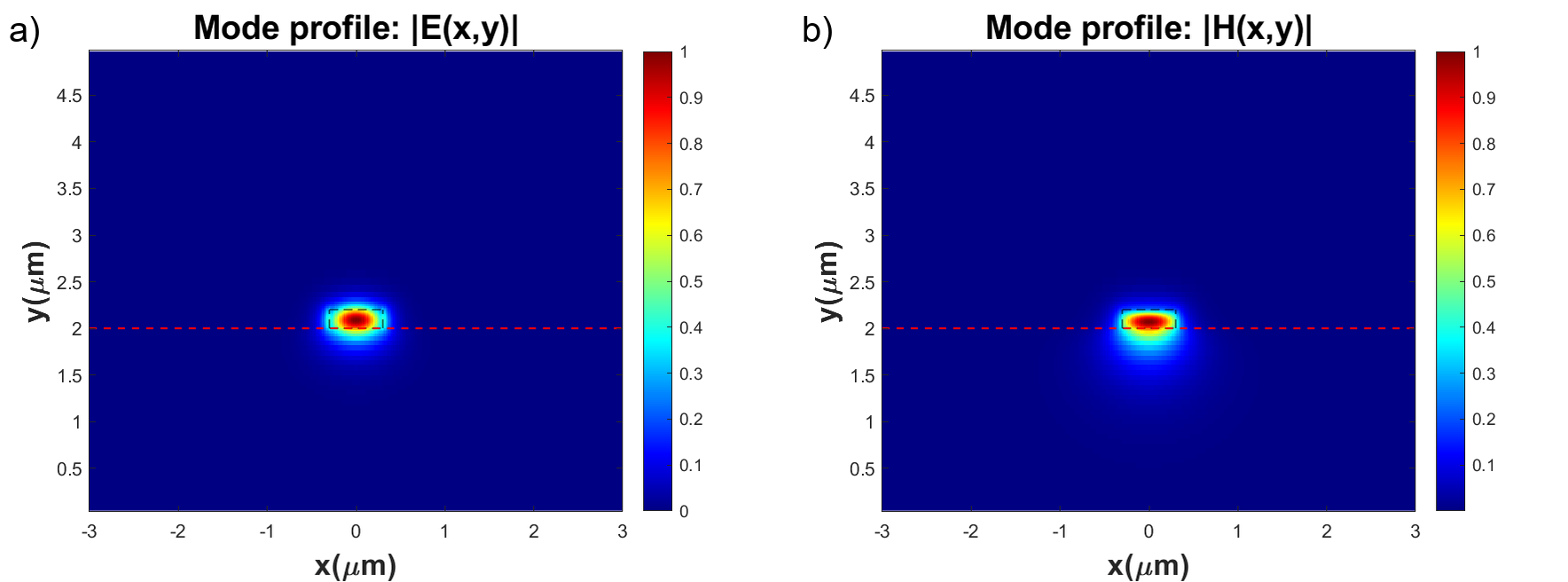}
\caption{a) Mode profile ($|\text{E(x,y)}|$) of $\text{TE}_{0}$ mode confined in 200 nm thick and 600 nm wide $\text{Si}_{3}\text{N}_{4}$ core waveguide on top of 2 $\mu\text{m}$ $\text{SiO}_{2}$ on Si substrate. The black dotted line shows the core waveguide, and the red dotted line shows the interface of $\text{SiO}_{2}$. b) Mode profile ($|\text{H(x,y)}|$) of $\text{TM}_{0}$ mode confined in 200 nm thick and 600 nm wide $\text{Si}_{3}\text{N}_{4}$ core waveguide on top of 2 $\mu\text{m}$ $\text{SiO}_{2}$ on Si substrate. The black dotted line shows the core waveguide, and the red dotted line shows the interface of $\text{SiO}_{2}$.}
\label{MP_array1}
\end{figure}

\begin{figure}[ht!]
\centering\includegraphics[width=8.5cm]{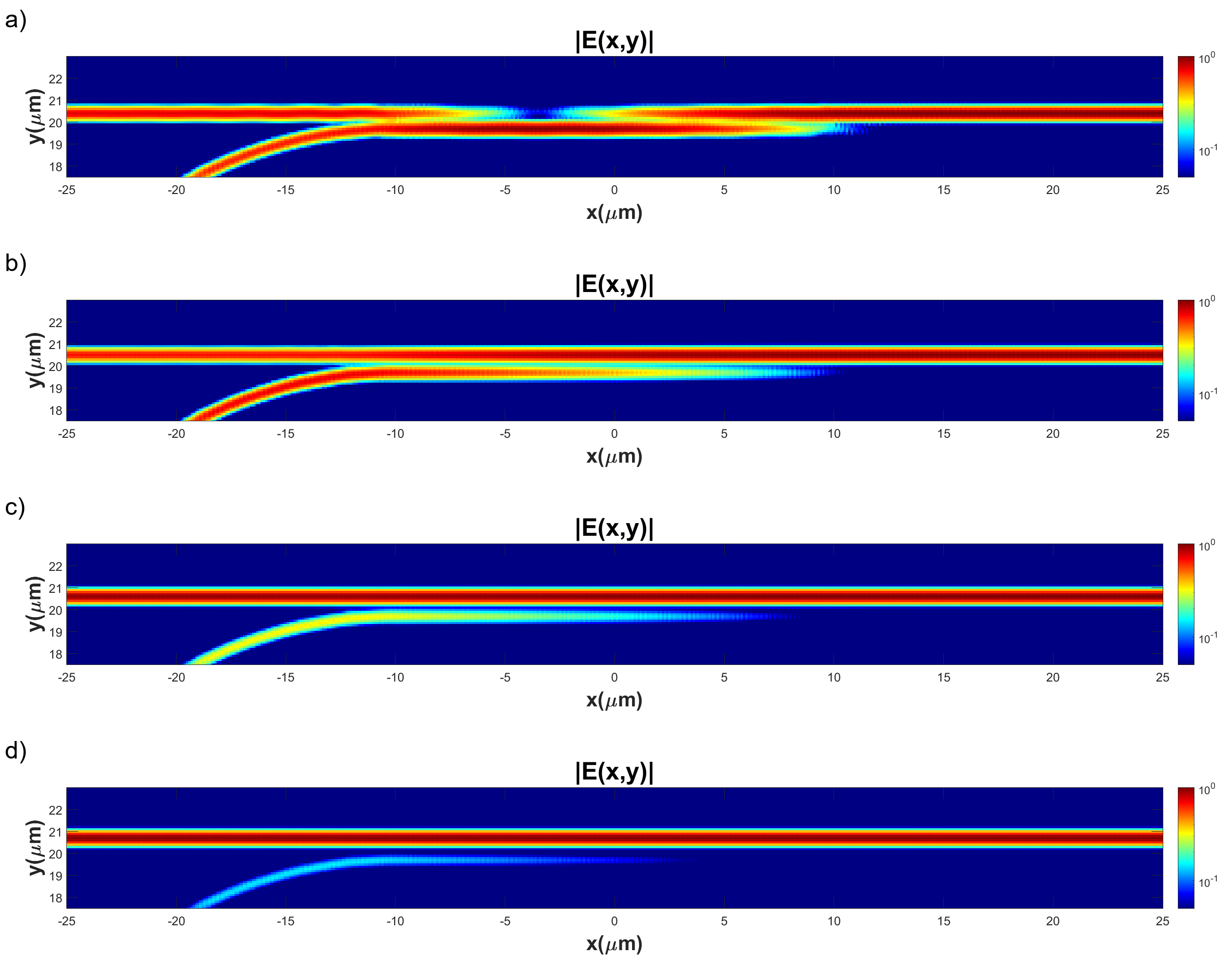}
\caption{($|\text{E(x,y)}|$) at z = 0 (center of 200 nm thick $\text{Si}_{3}\text{N}_{4}$ core waveguide). Propagation of light in $\text{TE}_{0}$ mode (from right to left) of the waveguide,  the gap between waveguide and resonator are a) 100 nm, b) 200 nm, c) 300 nm, d) 400 nm illustrating the intensity of light coupled from waveguide to resonator.}
\label{TE_array1}
\end{figure}

\begin{figure}[ht!]
\centering\includegraphics[width=8.5cm]{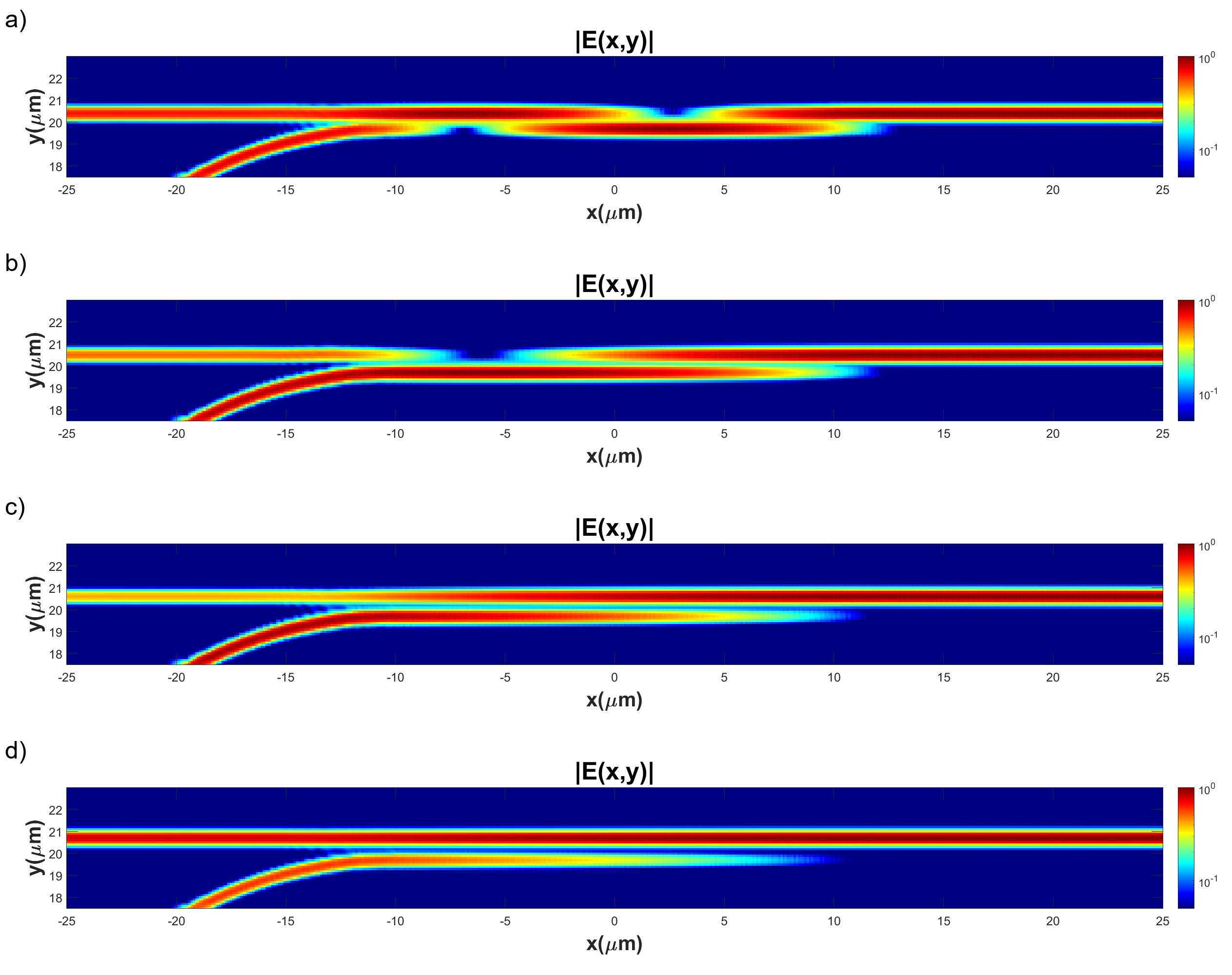}
\caption{$|\text{E(x,y)}|$ at z = 0 (center of 200 nm thick $\text{Si}_{3}\text{N}_{4}$ core waveguide). Propagation of light in $\text{TM}_{0}$ mode (from right to left) of the waveguide, the gap between waveguide and resonator are a) 100 nm, b) 200 nm, c) 300 nm, d) 400 nm illustrating the intensity of light coupled from waveguide to resonator.}
\label{TM_array1}
\end{figure}

\noindent We use 200 nm thick $\text{Si}_{3}\text{N}_{4}$ on 2 $\mu\text{m}$ $\text{SiO}_{2}$ clad on Si substrate that was readily available in the lab. We employ a racetrack-shaped resonator due to ease in fabricating the resonator and waveguide. The racetrack-shaped resonator has a longer coupling length or region for the bus-waveguide and resonator, which enables a relatively larger gap between them than compared to the circular-shaped (or ring) resonator for the same coupling constant; the schematics diagram is shown in Fig. \ref{GDS_image}. Effective indices of quasi-$\text{TE}_{0}$ (or TE$_{0}$), $\text{TM}_{0}$ and $\text{TE}_{1}$ modes confined in 200 nm thick $\text{Si}_{3}\text{N}_{4}$ varied width core waveguide on 2 $\mu m$ thick $\text{SiO}_{2}$ on Si substrate are estimated using finite-difference time-domain (FDTD) method as shown in Fig. \ref{neff_modes} while the group index of $\text{TE}_{0}$ and $\text{TM}_{0}$ modes are obtained from the slope of the dispersion curve $\text{n}_{\text{g}}=\text{n}_{\text{eff}}-\lambda\frac{\text{dn}_{\text{neff}}}{\text{d}\lambda}$ as shown in Fig. \ref{ng_all}. The design is chosen so that the length of the device is maximum and within the field view area of our homemade confocal microscope. Based on indices analysis as the plot shown in Fig. \ref{neff_modes}, we choose 600 nm width for our device, mode profiles of $\text{TE}_{0}$ and $\text{TM}_{0}$ modes are shown in Fig. \ref{MP_array1} and the coupling of light from waveguide to resonator for $\text{TE}_{0}$ and $\text{TM}_{0}$ modes are shown in Fig. \ref{TE_array1} and Fig. \ref{TM_array1} respectively. Since the $\text{TE}_{0}$ mode index is higher than the $\text{TM}_{0}$ mode, the light is more tightly confined within the waveguide for the $\text{TE}_{0}$ mode. This tighter confinement reduces the propagation losses. Consequently, in a resonator, the $\text{TE}_{0}$ mode exhibits a higher Q compared to the $\text{TM}_{0}$ mode. The higher Q-factor indicates lower energy dissipation per cycle \cite{M_ller_1977}, making the $\text{TE}_{0}$ mode more efficient for applications requiring high resonance stability and low loss, such as in high-performance optical sensors and devices \cite{Antonacci_2022, Vahala_2003}. The grating coupler parameters, such as grove pitch, width, and angle, are optimized by maximizing the coupling efficiency of a Gaussian source $\text{TEM}_{00}$ mode to $\text{TE}_{0}$ mode of waveguide around 770 nm in the FDTD method.


\section{Fabrication of Silicon Nitride Resonator}
\begin{figure}[ht!]
\centering\includegraphics[width=8.5cm]{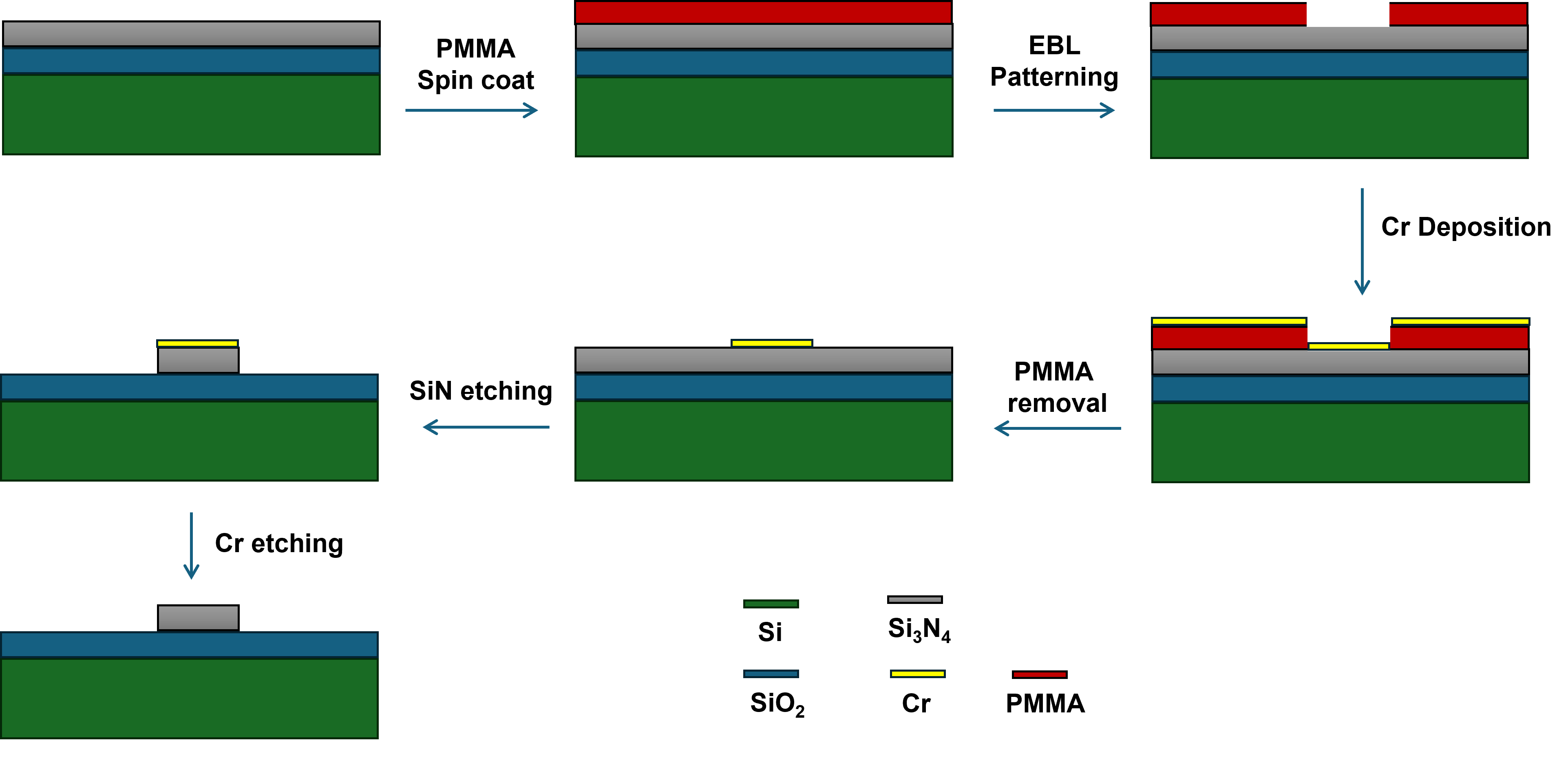}
\caption{Fabrication steps of $\text{Si}_{3}\text{N}_{4}$ resonator.}
\label{fabrication_steps}
\end{figure}

\noindent The fabrication procedure starts by cutting a small piece from a 3-inches wafer that has 200 nm of low-stress LPCVD silicon nitride film deposited on top of 2 $\mu m$ of $\text{SiO}_{2}$ on a Si substrate commercially available from Rouge Valley Microdevices. The sample undergoes a thorough cleaning process to remove any contaminants. Dehumidification is then performed by placing the sample on a hot plate at 180 $^\circ$C for 120 seconds to ensure any residual moisture is evaporated, followed by cooling it on a room temperature plate for one minute to stabilize it. Subsequently, polymethyl methacrylate (PMMA) 950 A4 is spin-coated onto the sample at 4000 rpm for 45 seconds to create a uniform resist layer. The sample is soft baked at 180$^\circ$C for 120 seconds and after soft baked and transferred to TESCAN MIRA3, an electron beam lithography tool. Resonator and waveguide patterns were written onto the resist layer with an exposure dose of 370 $\mu\text{C/cm}^2$, beam intensity 10, and an accelerating voltage of 30 kV. The pattern is developed in methyl isobutyl ketone (MIBK) and isopropyl alcohol (IPA) (1:3 ratio) solution for 60 seconds. The sample is dried using dry nitrogen gas to ensure no residual solvent is present. Next, a 40 nm thick layer of chromium (Cr) is deposited onto the sample using electron-beam physical vapor deposition (EB-PVD), forming a hard mask that will be used for the subsequent etching process. The resist-developed PMMA is then removed using acetone to leave the Cr mask intact on the sample. The $\text{Si}_{3}\text{N}_{4}$ layer is etched using plasma reactive ion etching (RIE) with a gas mixture of $\text{SF}_{6}$ at a flow rate of 50 sccm. The RIE process is carried out with a power of 150 W, a bias voltage of 500 V, and chamber pressure is set to 20 mTorr, selectively removing the $\text{Si}_{3}\text{N}_{4}$ in the areas not protected by the Cr mask. Following the $\text{Si}_{3}\text{N}_{4}$ etching, the Cr mask is removed using a chromium etchant, followed by thorough rinsing with deionized (DI) water and IPA. The flow chart of fabrication steps is shown in Fig. \ref{fabrication_steps}.

\begin{figure}[ht!]
\centering\includegraphics[width=8.5cm]{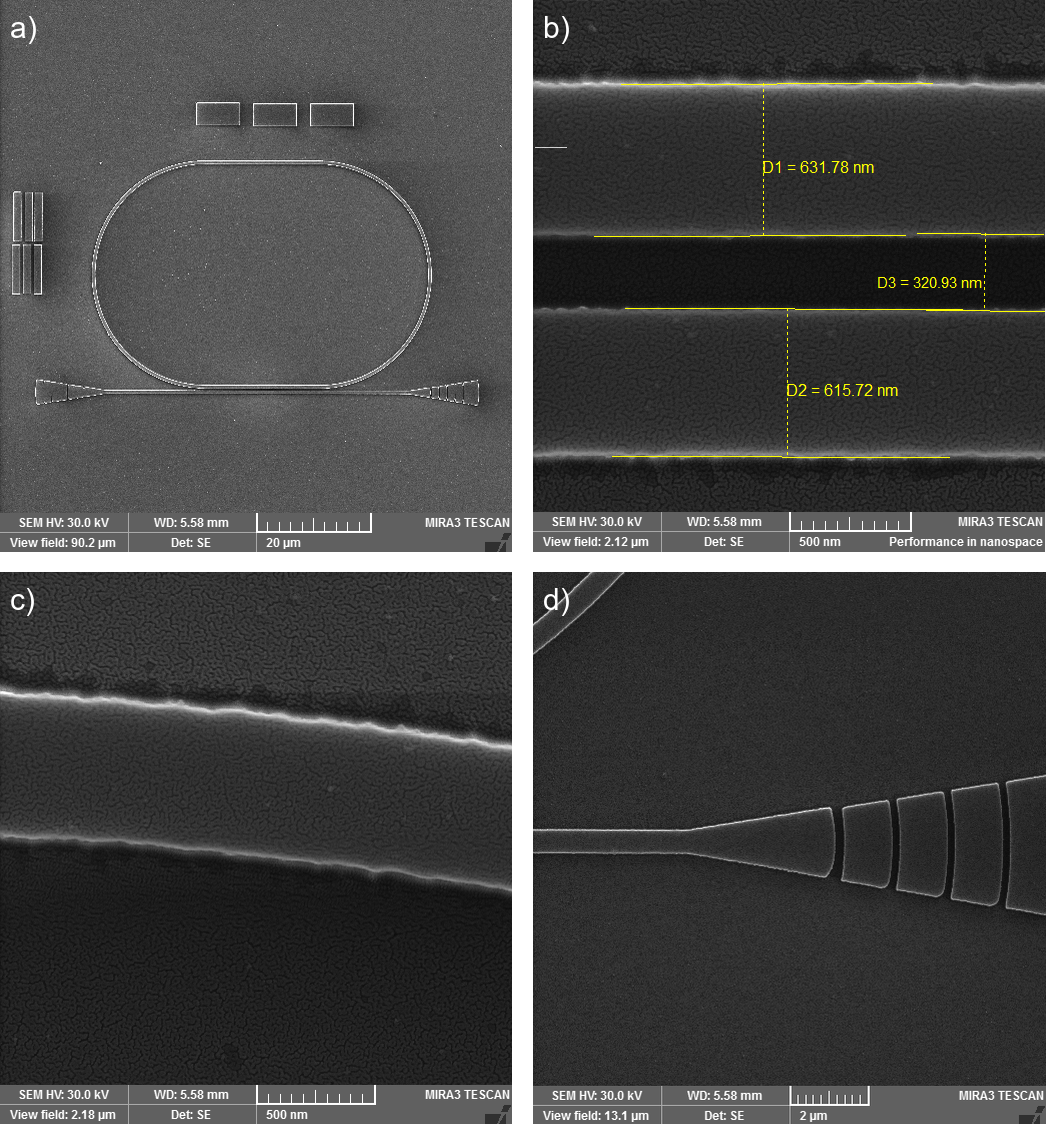}
\caption{Scanning electron microscope images of waveguide-coupled racetrack-shaped resonator (Device II): a) wider field view showing resonator and waveguide, extra rectangles on left and top of the resonator are used as markers, b) smaller field view at resonator and waveguide coupling region showing variation in widths of waveguide from desired 600 nm waveguide width, c) shows the roughness of waveguide edges, and d) waveguide-grating region of device.}
\label{sem_images}
\end{figure}

\section{Experimental Setup}
\begin{figure}[ht!]
\centering\includegraphics[width=8.5cm]{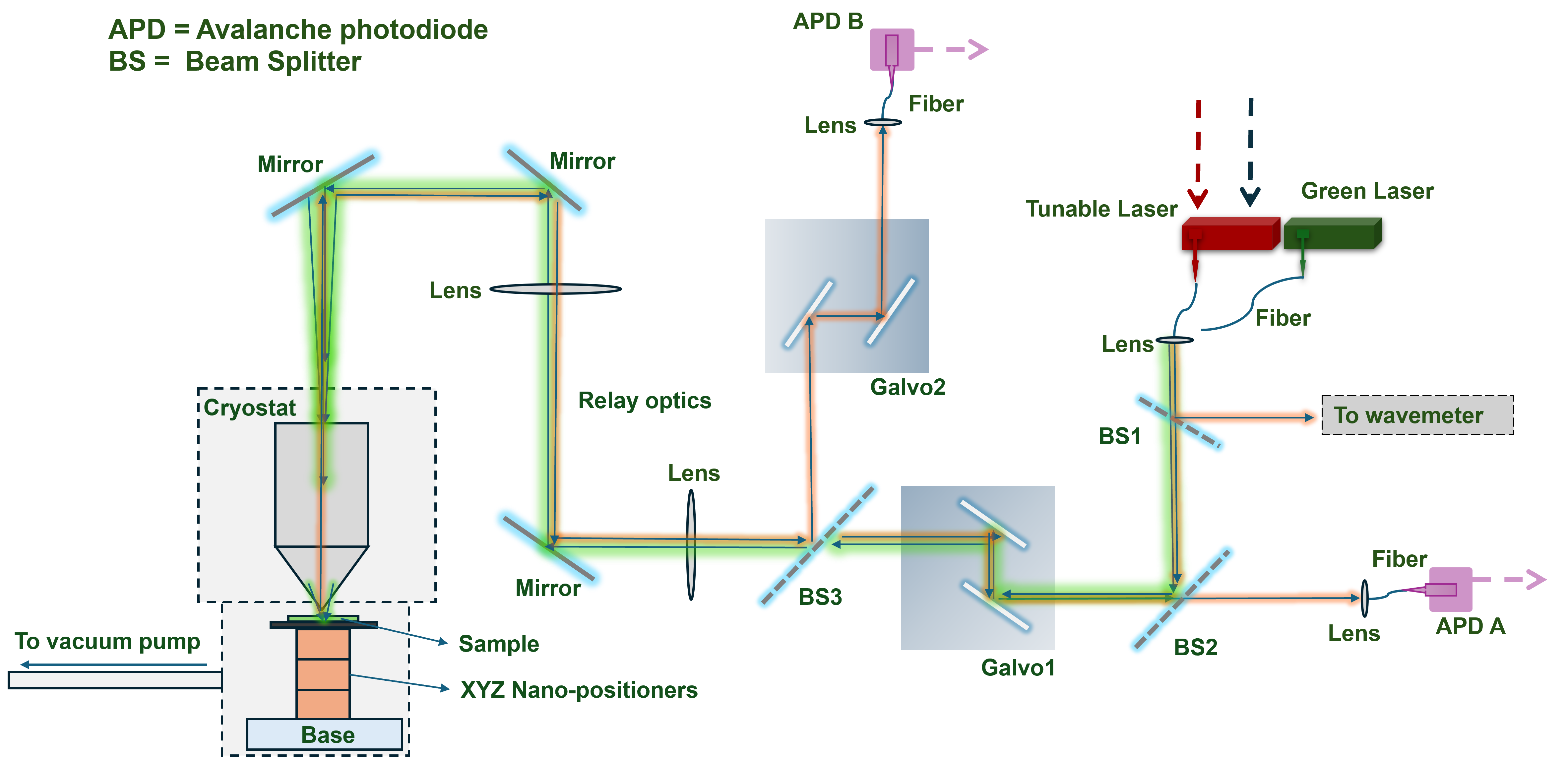}
\caption{Schematics diagram of a two-channel confocal laser scanning microscope for device characterization. The tunable laser (765-780 nm) is shown in red color, and the green laser (532 nm) is shown in green color. Two inputs are received by a tunable laser or its driver for laser operation (see Fig. \ref{measurement_setup}), where one input is from a computer to control the laser parameters and the second input is from a function generator for fine wavelength scanning or tuning. Electrical signals from APD A and B are sent to a computer via electrical cables. Fibers at APD A and B collection ends act as pinholes for respective channels. Two galvos, galvo1 and galvo2, perform the image scanning for ChA (photons collected in APD A) and ChB (photons collected in APD B), respectively. A closed cryostat system holds the sample on nanopositioners and is connected to a vacuum turbo-pump system.}
\label{confocal_laser_scanning_microscopy}
\end{figure}

\begin{figure}[ht!]
\centering\includegraphics[width=8.5cm]{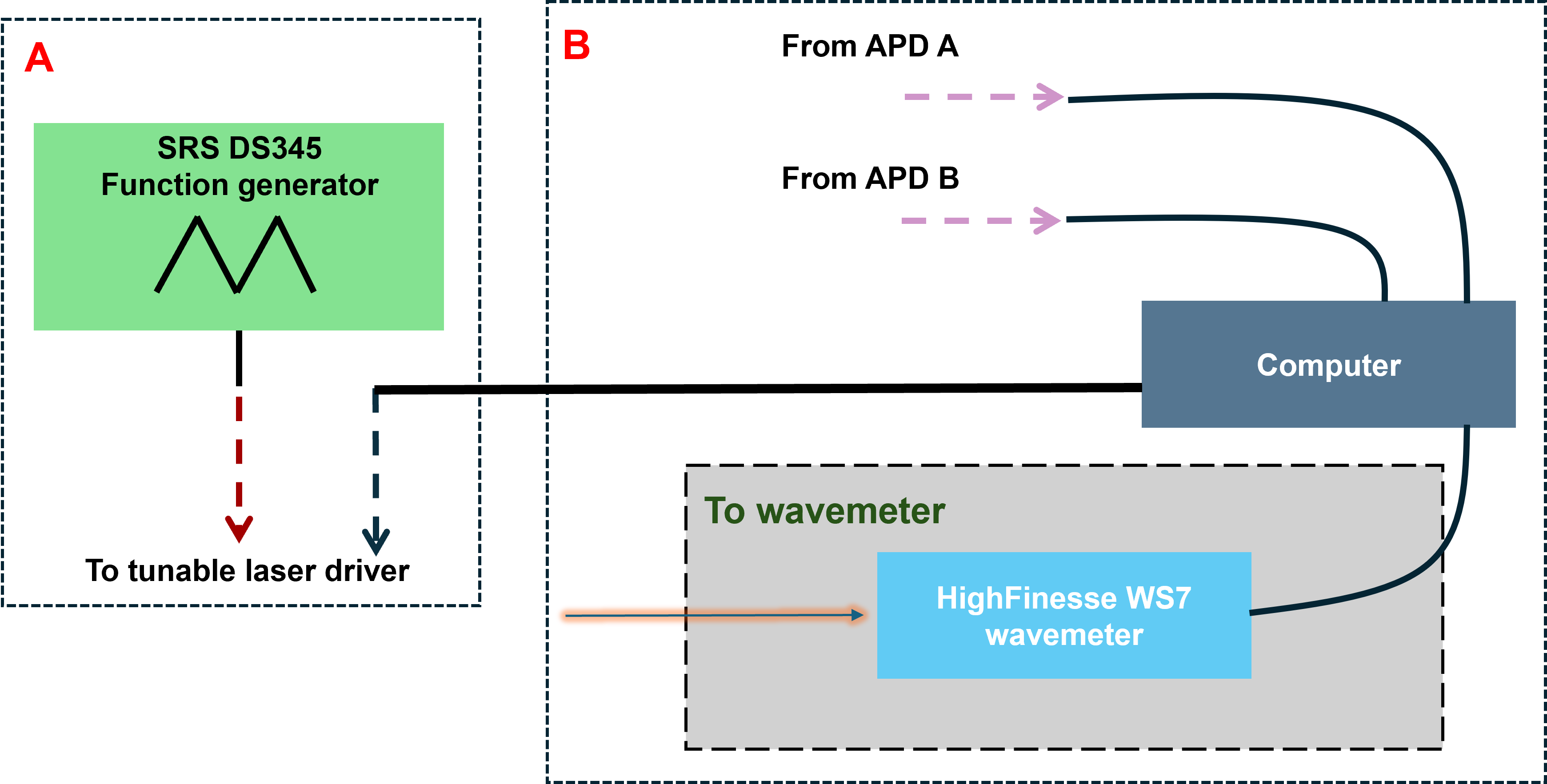}
\caption{Schematics diagram of the optical measurement and wavelength scanning setup. A shows the connection from a function generator to a tunable laser driver, and B shows the connection of wavemeter and APD signals to the computer for data recordings. The wavemeter collects the beam and uses an in-built interferometer to measure the source wavelength.}
\label{measurement_setup}
\end{figure}

\noindent 
We use homemade confocal microscopy with a high numerical aperture (NA) objective to demonstrate the mechanism of coupling light into and out of a photonic waveguide and the characterization of the resonator. We use the Attocube objective (LT-APO/532-RAMAN/0.82) with NA = 0.82 in the confocal microscope. The schematics diagram of a two-channel confocal microscope and measurement setup is shown in Fig. \ref{confocal_laser_scanning_microscopy} and \ref{measurement_setup}. The confocal setup uses two galvos mirrors (galvo1 and galvo2), three beam splitters (BS), one objective, two single-photon detectors, avalanche photodiodes (APD), and several lenses and mirrors. First, the collimated beam of laser is sent from the laser source (tunable laser or green laser) towards BS1 (90:10), 90\% of laser power is sent to the confocal setup, and 10\% of laser power is sent to the wavemeter (HighFinesse WS-7) (see Fig. \ref{measurement_setup}) for the measurement of the tunable laser source wavelength. The wavemeter is not employed when the green laser source is used. Next, the beam is directed towards galvo1 with BS2 (90:10). Galvo1 is used for channel A image scanning and sends the beam towards the objective. The objective is installed into a closed steel cryostat setup and sits in Janis ST-500 cryostat. The objective focuses the beam on the sample, which is placed under the focus and location of interest using XYZ Attocube nanopositioners. The sample reflects the light, couples out light via grating, or scatter light to the objective, and is sent towards the BS3 (50:50), and APD A and B collect these photons. Galvo2 is used for scanning channel B images. The area of the sample scannable by confocal setup is 100 $\mu$m $\times$ 100 $\mu$m, and both channels A and B collect photons from this area with high spatial and lateral resolution. The APD A and B electrical signals are sent to the National Instruments (NI) board installed in the computer for photon counting. The reading from the wavemeter is collected in a computer via an application. The tunable laser (New focus TLB700) source driver is connected to the computer, the commands are sent, and data is retrieved by a python code. To drive the laser source for fine-tuning ($\delta f \sim$ 100 GHz), the function generator (SRS DS345) sends the triangular shape function with frequency $\sim$ 0.05 Hz via BNC cable to the laser driver. Before the measurement procedure, the wavemeter is calibrated with a He-Ne laser. The linewidth of the laser is below 200 KHz, and the wavemeter has a resolution of 0.01 pm with a minimum $\sim$1 ms sampling time. Sampling rate $\sim$ 15 ms is used in wavelength measurements, and laser wavelength tuning speed $\sim $ 0.01-0.001 nm/s is kept in the finding of the resonances and Q-factor measurement. The photon counts from APDs are collected with a sampling rate of $\sim$ 17 Hz. Typically, $\sim$ 1-10 $\mu$W laser power is used in the demonstration experiment. No spectral optical filters were used in the confocal setup, but they can be used in setup for specific applications.

\section{Results and Discussions}
\subsection{Device characterization}
\begin{figure}[ht!]
\centering\includegraphics[width=8.5cm]{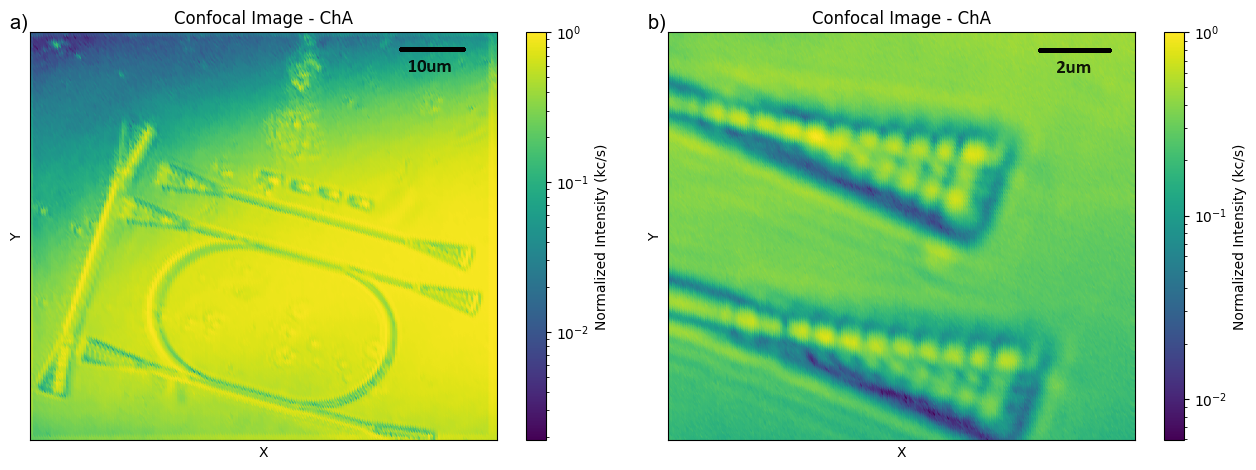}
\caption{Device I: a) ChA (reflected photons) confocal image obtained with 532 nm laser shows the resonator device coupled to waveguide (below and above of resonator) with gratings on each side (left and right), and additional waveguides for experimental investigation. b) ChA (reflected photons) confocal image zoomed-in at the grating region.}
\label{array1}
\end{figure}

\begin{figure}[ht!]
\centering\includegraphics[width=8.5cm]{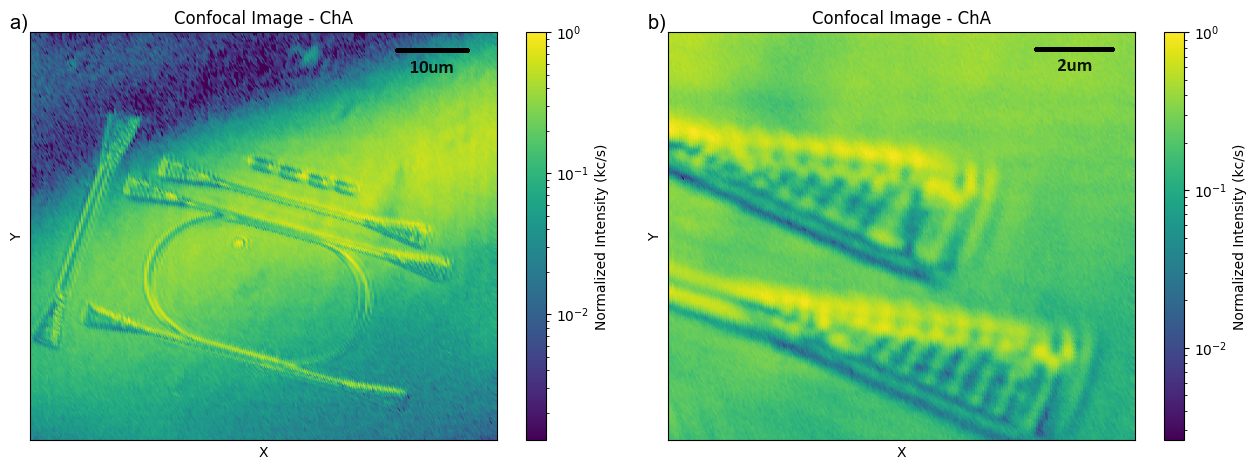}
\caption{Device I: a) ChA (reflected photons) confocal image obtained with 770 nm laser shows the left and right gratings, resonator coupled to waveguide (below and above of resonator), and extra waveguides. b) ChA (reflected photons) confocal image of zoomed-in at the grating region (on the right of the resonator).}
\label{array2}
\end{figure}

\begin{figure}[ht!]
\centering\includegraphics[width=8.5cm]{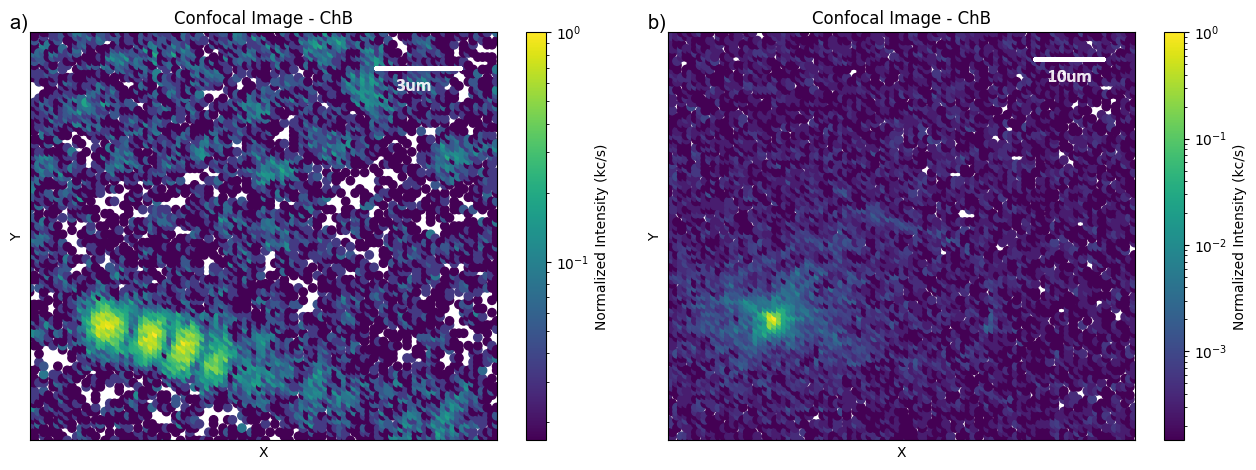}
\caption{Device I: a) ChB (collected photons) confocal image zoomed-in at the output grating region when the grating on the left of the waveguide coupled resonator is excited at a non-resonant wavelength in ChA, the excitation location at the grating in ChA is optimized to maximize the photons counts or signal in ChB. b) ChB (collected photons) confocal image at resonant wavelength 772.435 nm illustrates illumination (photons radiated out or scattered from resonator) of resonator due to loss of photons while propagating in the resonator.}
\label{array3}
\end{figure}

\begin{figure}[ht!]
\centering\includegraphics[width=8.5cm]{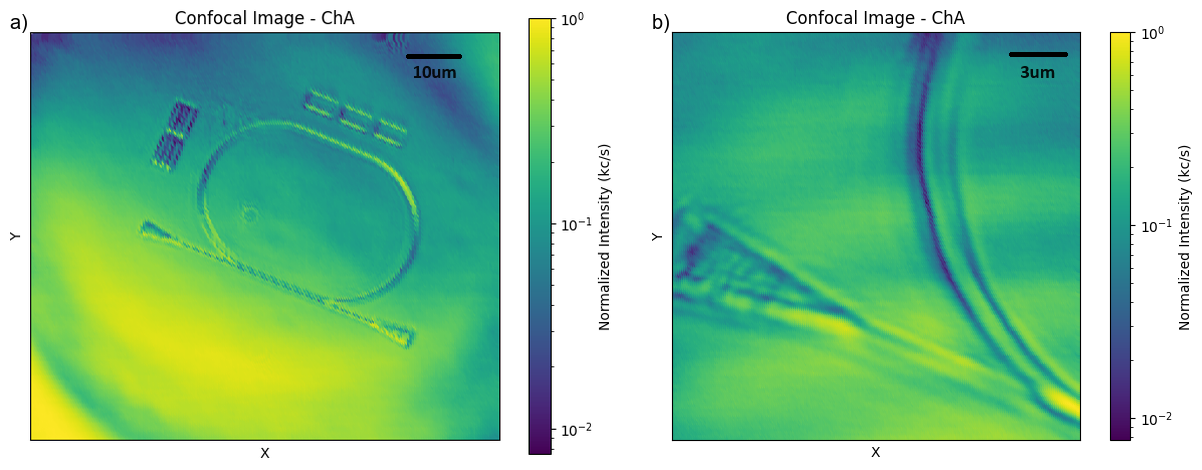}
\caption{Device II: a) ChA (reflected photons) confocal image obtained with 770 nm laser shows left and right gratings, waveguide coupled resonator. b) ChA (reflected photons) confocal image of zoomed-in at the grating region (the one on the left side of the resonator).}
\label{array4}
\end{figure}

\begin{figure}[ht!]
\centering\includegraphics[width=8.5cm]{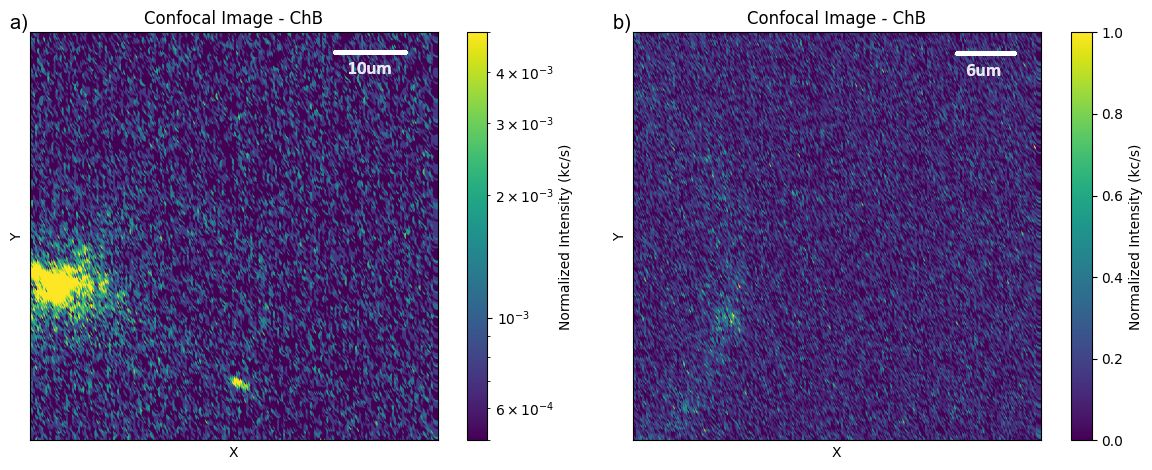}
\caption{Device II: a) ChB (collected photons) confocal image of at resonant wavelength 772.667 nm illustrates the illumination of resonator due to loss of photons while propagating in the resonator. b) ChB (collected photons) confocal zoomed-in image of a small portion of the illuminated resonator, as shown in part (a) of this figure.}
\label{array5}
\end{figure}

\noindent 
One channel (A) of the confocal microscope is used for excitation, and the second channel (B) collects the transmission through waveguide via gratings as demonstrated in Fig. \ref{array1}, \ref{array2}, \ref{array3}, 
 \ref{array4} and \ref{array5}.
 
\noindent
First, a high-resolution confocal image of the resonator device is acquired using a 532 nm laser from ChA, which captures the reflected photons from the sample to APD A. The experimental setup is depicted in Fig. \ref{confocal_laser_scanning_microscopy}, and the obtained image is shown in Fig. \ref{array1}. Subsequently, another confocal image is obtained from ChA using a tunable laser around 770 nm; the obtained image is shown in Fig. \ref{array2}. Grating \#1 (selective either the left or right) is excited through ChA with 770 nm laser, and ChB collects the photons from the sample to APD B, and the corresponding images are captured (see Fig. \ref{array3}). Optimization of the laser excitation position on Grating \#1 in ChA is performed to maximize photon counts from Grating \#2 in ChB. In the ChB image, the location of the maximum photon counts at the output grating (see Fig. \ref{array3} (a)) is selected for the measurement. Device I was used in the early stage of the experimental setup investigations, and later, device II was used for Q-factor and sensing measurements. Throughout these processes, images acquired are shown in Fig. \ref{array3}, \ref{array4}, and \ref{array5}. Wavelength scanning is conducted to locate the resonant wavelength of the resonator. Resonances, Q-factor, and finesse are determined by measuring (APD A: reflection at grating excitation in ChA)/(APD B: collection at output grating in ChB) ratio as the output signal. This ratio is taken as the output signal to prevent any sharp fluctuation in the signal due to input laser power fluctuations during wavelength scanning and any disturbances to the setup. At resonance, the light couples into the resonator, and the photons radiate out of the resonator or scatter out from the resonator (or in other words, the resonator is illuminated at the resonance), which are collected in ChB, as illustrated in Fig. \ref{array3}(b) and \ref{array5}. The resonator illumination at its resonant wavelength in ChB is compelling evidence of the excitation of resonance. Note that these photons are collected into a fiber (ChB) from a small portion of a resonator with a high signal-to-noise ratio, which is an advantage compared to the standard characterization method.

\begin{figure}[ht!]
\centering\includegraphics[width=8.5cm]{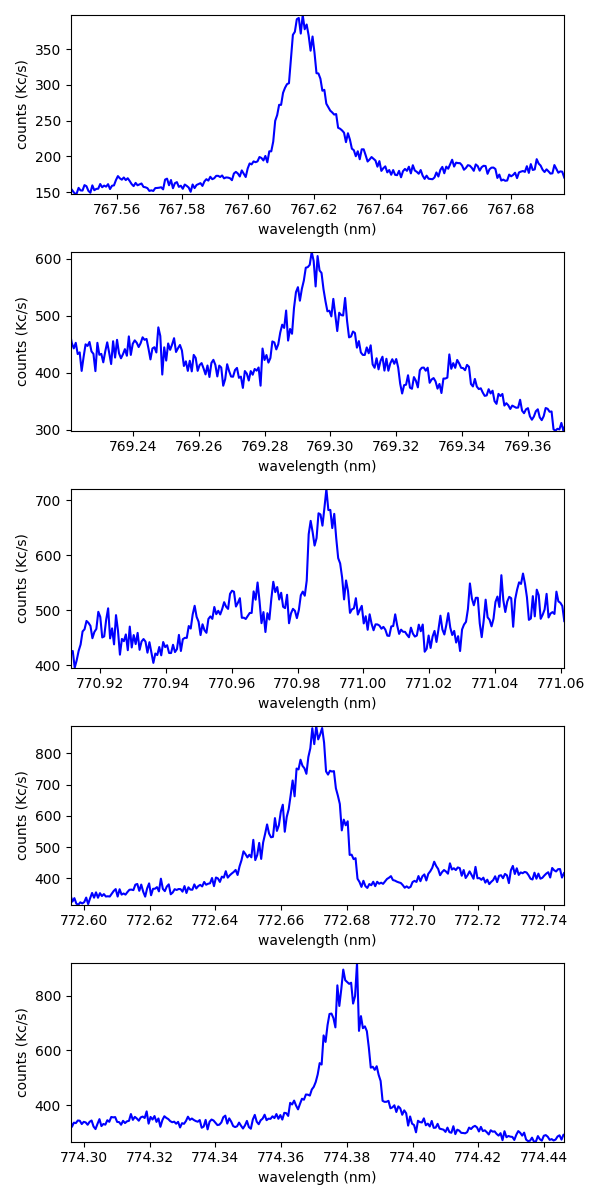}
\caption{Device II: Five longitudinal resonance modes measured with tunable laser (coarse wavelength scanning $\sim$ 0.01 nm/s), counts (Kc/s) on the y-axis is A(reflection)/B(collection) ratio signal obtained in the measurement at 299.33 K temperature.}
\label{resonances_TE0_measured_values}
\end{figure}

\begin{figure}[ht!]
\centering\includegraphics[width=8.5cm]{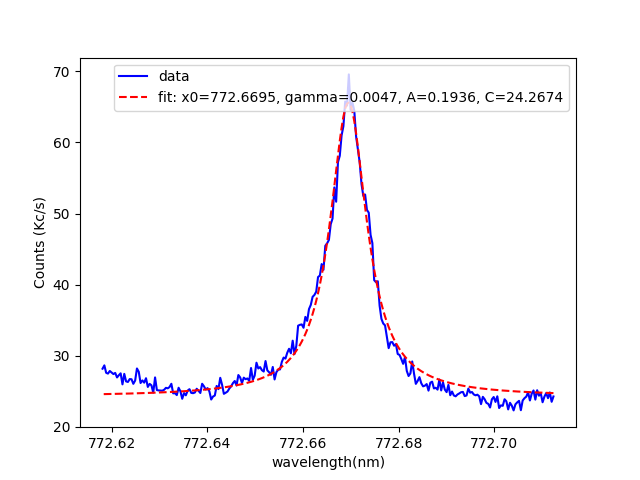}
\caption{Device II: Resonance measured at 772.669 nm tunable laser (fine wavelength scanning $<$ 0.001 nm/s), counts (Kc/s) on the y-axis is APDs A(reflection)/B(collection) ratio signal obtained in the measurement. The curve is fitted with Lorentzian function, $C + \frac{A\gamma}{(x-x_{0})^{2} + \gamma^{2}}$.}
\label{resonance_Q_factor}
\end{figure}

\noindent Systematic scanning of resonance is performed to measure Q-factor and finesse. Fig. \ref{resonances_TE0_measured_values} shows the measured five longitudinal resonances with a tunable laser at 299.33 K. Loaded Q-factor ($Q_{load}$) of resonance is estimated around 772.67 nm by Lorentzian fit as shown in Fig. \ref{resonance_Q_factor} and is $\sim 8.2\pm 0.17\times10^{4}$. Intrinsic Q-factor ($Q_{in}$) of the resonator is calculated using $Q_{in}=2Q_{load}/(1+\sqrt{T_{0}})$, where $Q_{load}$ is loaded Q-factor = $\lambda/\delta \lambda$ found from the measurement and $T_{0}$ is fraction of transmitted power at resonance wavelength, and is $\sim$ 1 $\times$ 10$^{5}$. The propagation loss $\alpha$ (in cm$^{-1}$) is calculated by $\alpha=2\pi n_{g}/(Q_{in}\lambda)$ and is 0.84 dB/cm. 

\subsection{Air density sensing}

\begin{figure}[ht!]
\centering\includegraphics[width=8.5cm]{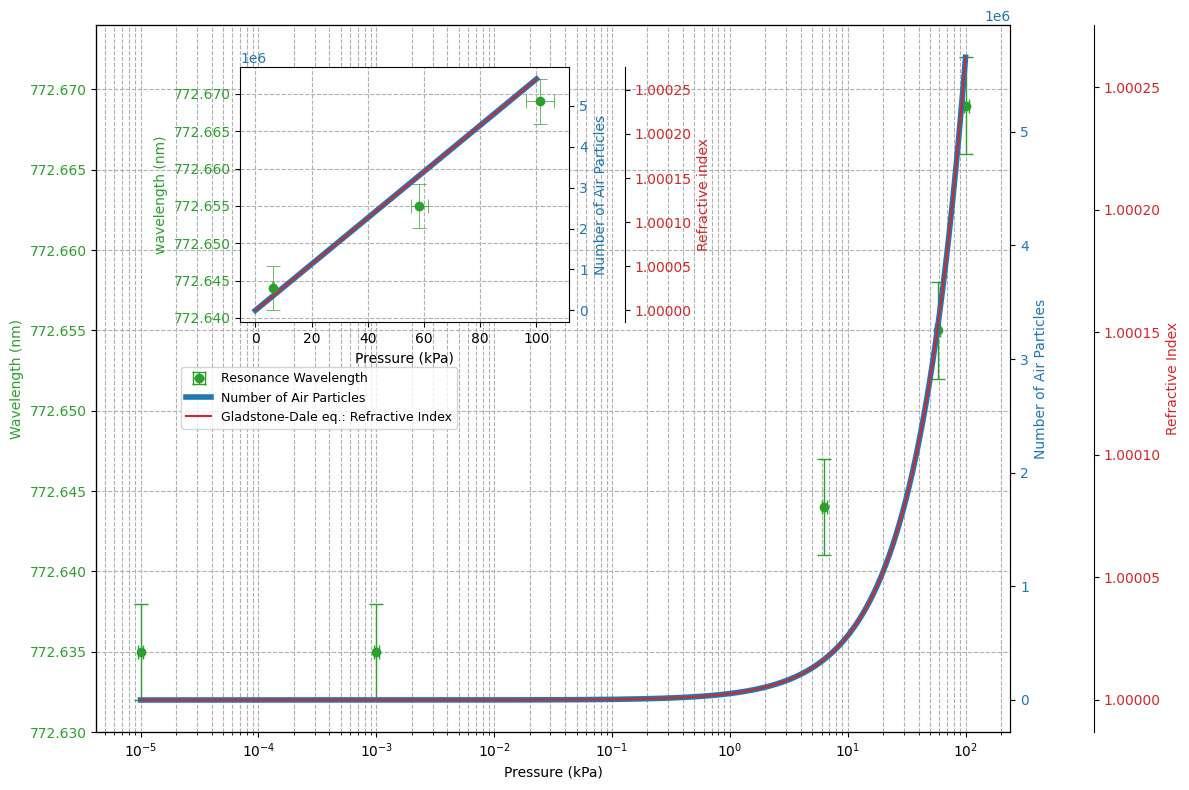}
\caption{Study of resonance shift with varied air pressure. The resonance wavelength on the y-axis and pressure on the x-axis is experimental data, and the number of air particles and refractive index of air curves are empirical. The inset plot shows the three data points for air pressures above 1 kPa.}
\label{peak_shift}
\end{figure}

\noindent To perform air density sensing measurements, a vacuum pump was employed to control the air pressure inside the cryostat as shown in the experimental setup Fig. \ref{confocal_laser_scanning_microscopy}. This setup allowed for systematically manipulating air density within the controlled environment. Resonance measurements were performed for device II in response to varying air pressure. The measurement data plotted in Fig \ref{peak_shift} illustrates the resonator response characteristics under different air pressures or densities at 299.33 K. The pressure values are measured with a pressure gauge, which has a 5\% error in measurement, so we include this error in our pressure recordings. Since the FWHM of resonance is $\sim 0.0094$ nm, an error in resonance wavelength can be safely taken as 1/3 of FWHM for high credibility in detection measurements. Therefore, the error in wavelength values equals $\pm$ 0.0031 nm. The plot also includes the number of air particles in sensing volume and the refractive index of air relation with air pressure inside the sample chamber, which are evaluated in the following manner. For the $\text{TE}_{0}$ mode confined in 200 nm thick and 600 nm wide $\text{Si}_{3}\text{N}_{4}$ core waveguide, using FDTD analysis we estimate,

$$\max[\text{n}^{2}\text{(x,y)}|\text{E(x,y)}|^{2}]_\text{mode} = \text{4.000}$$

$$\iint_\text{mode} \text{n}^{2}\text{(x,y)}|\text{E(x,y)}|^{2} \,\text{dx}\,\text{dy} = \text{2.854}\times\text{10}^{-13}\,\,\text{m}^{2},$$ and $$\iint_\text{air} \text{n}^{2}\text{(x,y)}|\text{E(x,y)}|^{2} \,\text{dx}\,\text{dy} = \text{5.669}\times\text{10}^{-15}\,\,\text{m}^{2},$$
therefore,
$$\text{A}_{\text{eff,air}}=\text{1.417}\times\text{10}^{-15}\,\,\text{m}^{2},$$
$$\text{V}_{\text{sensing}}=\text{L}_{\text{resonator}}\times\text{A}_{\text{eff,air}}\sim\text{233.8}\times\text{10}^{-21}\,\,\text{m}^{3},$$
$$\text{Number of Particles in sensing volume}= \frac{\text{PV}_{\text{sensing}}}{\text{K}_{\text{B}}\text{T}},$$
$$\text{n(refractive index of air)}=\text{1 + K}\left(\frac{\text{PM}}{\text{RT}}\right)$$
where K = 0.000225, Gladstone-Dale constant for air ($\text{m}^{3}\text{/kg}$), M = 0.029, Molar mass of air (kg/mol), R = 8.314, Universal gas constant (J/(mol K)). The  plot shows the sensitivity level $\sim \text{10}^{6}$ air particles. Note that the thermo-optic coefficients ($dn/dT$) of $Si_{3}N_{4}$ and $SiO_{2}$ are 2.5 $\times$ $10^{-5}$ RIU/$^\circ$C and 0.95 $\times$ $10^{-5}$ RIU/$^\circ$C, respectively \cite{Arbabi2013}. The temperature recorded during the whole measurement was 299.33 K. Any slight change in temperature below 0.01 K produces two orders of magnitude more minor changes in the refractive index of the materials than that produced by air pressure in our sensing measurements.

\noindent 
In various applications, the sensitivity of optical resonators to air pressure may need to be either minimized or maximized, depending on the specific requirements. The interaction between the evanescent field of the resonator mode and the surrounding air contributes significantly to the resonator's response to pressure changes at a constant temperature. This interaction is largely determined by the mode overlap with the air outside the core waveguide. To achieve either air-insensitive or optimally sensitive resonators, the geometric parameters of the core waveguide, particularly its thickness and width, can be strategically manipulated. This tailoring of the waveguide dimensions allows for control over the extent of the evanescent field and, consequently, the magnitude of the wavelength shift in response to pressure variations. Such precise engineering of resonator sensitivity can be crucial in applications involving the coupling of nanodiamonds \cite{Alkahtani_2019}, quantum dots \cite{Jonas_2022, Tang_2022}, upconversion particles \cite{Rajil_2022, Esmaeili, Esmaeili2024, Esmaeili2022} to optical resonator under different-pressure conditions, where maintaining a stable resonance or achieving high sensitivity to minute pressure changes can significantly impact the device performance and their applications. This approach enables the design of resonators with characteristics specifically suited to their intended applications, whether they require minimal response to ambient pressure fluctuations or enhanced sensitivity for particle-based sensing.

\section{Conclusion}
\noindent 
We presented the design, simulations, and fabrication of racetrack-shaped $\text{Si}_{3}\text{N}_{4}$ resonator and demonstrated its characterization using confocal laser scanning microscopy. We presented numerical estimations of indices of modes in different configurations of the core waveguide and achieved loaded Q-factor $\sim \text{8.2}\pm \text{0.17} \times \text{10}^{4}$ and finesse $\sim \text{180}\pm \text{3.5}$. Using this method, we demonstrated an application for air density sensing measurement. The results provide a relationship between air pressure and the measured resonance shifts, offering insights into the sensor's performance, sensitivity, and potential applications in air quality monitoring or environmental sensing. This sensor could be implemented to detect the relatively small number of bio-molecules present in air, gas, or liquids in the future bio-sensor.

\noindent While we have used the confocal microscopy method to characterize photonic resonators at 770 nm, it can be expanded across many disciplines in determining quantitative information of photonic devices applicable in fields such as biomedical sciences, quantum computing, electro-optic devices, optical sensors \cite{Lukin2020-dn, Rugar2020-ge}. Although we have limited our demonstration of the resonator to $\text{TE}_{0}$ mode only, both $\text{TE}_{0}$ and $\text{TM}_{0}$ modes of a waveguide play distinct roles in the interaction with particles or molecules positioned on its surface. In the $\text{TE}_{0}$ mode, the electric field is entirely perpendicular to the waveguide's cross-sectional plane, while in the $\text{TM}_{0}$ mode, the magnetic field is perpendicular to this plane \cite{Ahmed_2015}. $\text{TM}_{0}$ mode is less commonly exploited in sensing applications but finds utility in scenarios requiring distinct interactions based on magnetic properties or current induction in nearby materials.

\section{Funding}
\noindent
A.H. is supported by Herman F. Heep and Minnie Belle Heep Texas A\&M University Endowed Fund held/administered by the Texas A\&M Foundation. We would like to thank the Robert A. Welch Foundation (grants A-1261 and A-1547), the DARPA PhENOM program, the Air Force Office of Scientific Research (Award No. FA9550-20-10366), and the National Science Foundation (Grant No. PHY-2013771). This material is also based upon work supported by the U.S. Department of Energy, Office of Science, Office of Biological and Environmental Research under Award Number DE-SC-0023103, SUB-2023-10388, and FWPERW7011.

\section{Acknowledgments}
\noindent
Photonic chips were fabricated at the Aggiefab facility of Texas A\&M University. Portions of this research were conducted with the advanced computing resources provided by Texas A\&M High Performance Research Computing.
 
\section{Disclosures}
\noindent
The authors declare that they have no known competing financial interests or personal relationships that could have appeared to influence the work reported in this paper.

\section{Author contribution}
\noindent
M.K. and A.H. performed the experiment and wrote the manuscript. L.L. was involved in an early-stage experiment. M.K., A.H., and L.L. optimized the fabrication recipe. M.K. performed the simulation and design of devices. M.K., P.R.H., Z.Y., A.S., and M.O.S. conceived the idea.

\section{Data Availability}
\noindent
Data underlying the results presented in this paper are not publicly available at this time but can be obtained from the authors upon reasonable request.

\bibliography{apssamp}

\end{document}